\vspace{0.2cm}
\pdfoutput=1
\documentclass[a4paper,11pt]{article}
\usepackage{amsmath,amssymb,amsfonts}
\usepackage[multiple,stable]{footmisc}
\usepackage[amsmath,amsthm,thmmarks]{ntheorem}
\allowdisplaybreaks[4]
\usepackage[noconfig]{refstyle}
\usepackage[dvipsnames]{xcolor}
\usepackage[hyperfootnotes=false, linktocpage=true, colorlinks, citecolor=blue, linkcolor=blue, urlcolor=Maroon]{hyperref}
\usepackage{array,tabularx,booktabs,geometry,subfigure,graphicx,longtable}
\usepackage[bf]{caption}
\usepackage{cite}
\usepackage{tikz}
\usetikzlibrary{shapes,arrows}
\tikzstyle{block} = [rectangle, draw, text width=7em, text centered, rounded corners, minimum height=3em]

\numberwithin{equation}{section}

\newcommand{\be}{\begin{equation}}
\newcommand{\ee}{\end{equation}}
\newcommand{\bea}{\begin{equation}\begin{aligned}}	
\newcommand{\eea}{\end{aligned}\end{equation}}		

\newcommand{\iddots}{\mathinner{\mkern2mu\raise1pt\hbox{.}\mkern2mu \raise4pt\hbox{.}\mkern2mu\raise7pt\hbox{.}\mkern1mu}}

\providecommand{\id}{\leavevmode\hbox{\small$\mathrm{1}$\kern-3.8pt\normalsize$\mathrm{1}$}}
\def\fnote#1#2{\begingroup\def\thefootnote{#1}\footnote{#2}
     \addtocounter{footnote}{-1}\endgroup}

\begin{document}

\vspace{1cm}

\title{
       \vskip 40pt
       {\LARGE \bf Vanishing conditions for higher order\\  couplings in heterotic theories}}

\vspace{2cm}

\author{James Gray}
\date{}
\maketitle
\begin{center} {\small {\it Physics Department, Robeson Hall, Virginia Tech,\\ Blacksburg, VA 24061, U.S.A.}}\\
\fnote{}{jamesgray@vt.edu}
\end{center}

\begin{abstract}
\noindent
For compactifications of heterotic string theory, we elucidate simple cohomological conditions that lead to the vanishing of superpotential n-point couplings for all n. These results generalize  some vanishing theorems for Yukawa couplings that have previously appeared in the literature to all higher orders. In some cases, these results are enough to show that certain fields do not appear in the perturbative superpotential at all. We illustrate our discussion with a number of concrete examples. In some cases, our results can be confirmed by showing that symmetries indeed forbid the couplings that vanish. In many, however, no such symmetries are known to exist and, as such, the infinite sets of vanishing couplings that are found are surprising from a four-dimensional perspective. 
\end{abstract}

\thispagestyle{empty}
\setcounter{page}{0}
\newpage

\section{Introduction}

Perturbative superpotential Yukawa couplings are a very well studied topic within the context of compactifications of the heterotic string on smooth Calabi-Yau threefolds \cite{Strominger:1985ks,Greene:1986bm,Greene:1986jb,Candelas:1987se,Distler:1987gg,Distler:1987ee,Greene:1987xh,Candelas:1987rx,Candelas:1990pi,Distler:1995bc,Berglund:1995yu,Braun:2006me,Bouchard:2006dn,Anderson:2010vdj,Anderson:2010tc,Buchbinder:2014sya,Blesneag:2015pvz,Blesneag:2016yag,McOrist:2016cfl,Ashmore:2018ybe,Gray:2019tzn,Anderson:2021unr,Anderson:2022kgk,Ibarra:2024hfm} (see \cite{Butbaia:2024tje,Constantin:2024yxh} for some recent progress on physically normalized Yukawa couplings). One of the more interesting structures that has been seen within this context is that such couplings frequently vanish in a manner that is somewhat surprising from a four dimensional field theory perspective (see \cite{Braun:2006me,Bouchard:2006dn,Anderson:2010tc,Buchbinder:2014sya,Blesneag:2015pvz,Blesneag:2016yag,Gray:2019tzn,Anderson:2021unr,Anderson:2022kgk} for some examples). In particular, in the constructions that appear in the literature, it is frequently the case that perturbative superpotential Yukawa couplings are zero despite the fact that no known symmetry enforces this structure upon the theory. These vanishings of trilinear couplings seem to arise from a plethora of different geometric origins, from the fact that the three-fold can be embedded into an ambient space, to the existence of fibration structure in the geometry and more besides \cite{Braun:2006me,Bouchard:2006dn,Blesneag:2015pvz,Blesneag:2016yag,Gray:2019tzn,Anderson:2021unr,Anderson:2022kgk}.

In physics we are interest in couplings beyond cubic, renormalizable, interactions. A natural question is then whether the phenomenon of vanishing couplings extends to these higher order interactions? We stress here that we are not talking about terms which are higher order in $\alpha'$ or the string coupling. The superpotential is computed exactly in these expansions due to non-renormalization results \cite{Green:1987mn}. Rather we are referring to superpotential contributions that are higher order in the matter fields \cite{Berglund:1995yu,Anderson:2011ty}\footnote{It should be noted that the matter fields appearing in a four dimensional heterotic theory, as conventionally parameterized, are not canonically normalized. As such no suppression scale appears in the superpotential for higher order terms such as those being discussed here. Such a scale (the compactification scale) would be reintroduced upon canonical normalization of the fields, leading to the usual suppression of higher order terms at low energies, without violating the standard non-renormalization results.}. In some sense, one example of vanishing of these higher order couplings is already included in the Yukawa coupling analysis itself. Vanishings of those interactions are frequently shown to hold for all values of the complex structure and/or bundle moduli. As such one can view these results as describing the vanishing of an infinite number of higher order couplings involving three matter fields and an arbitrary number of moduli. Indeed, there is a variety of other pieces of evidence that we can expect the phenomenon of vanishing couplings to go beyond cubic order in heterotic compactifications. For example, in many, although not all \cite{Thomas:1999tq,Anderson:2010mh,Anderson:2011ty}, commonly used bundle constructions one obtains a good heterotic background for an entire moduli space of defining data of the manifold and bundle.  In such an instance, the combined complex structure and bundle moduli space should be unobstructed, implying, for example, that all perturbative couplings between bundle moduli, of any order, should vanish. Since the bundle moduli are singlets, there is frequently no obvious symmetry in the four dimensional theory that would forbid couplings between these degrees of freedom. Another example is the fact that one frequently obtains heterotic models with massless vector like matter, say ${\bf 10}$'s of $SO(10)$, at generic points in moduli space. There is again often no known symmetry forbidding mass terms for such fields, and yet all couplings between two ${\bf 10}$'s and arbitrary numbers of moduli must vanish to be consistent with the massless spectrum computation. Note that supersymmetry is of course not an explanation for this phenomenon as any appropriate holomorphic couplings are allowed within the superpotential.

In this paper we will present a simple analysis which provides one reason why these couplings, and others between chiral multiplets for example, can vanish to all orders in the fields. The spirit of this work is similar to that of the description of vanishing Yukawa couplings mentioned above. It centers around ancillary geometric structures in the constructions such as the existence of descriptions of the compactifications involving an ambient space. Our results describe how a (small) finite number of simple cohomology constraints in the description of a heterotic compactification can lead to these infinite numbers of vanishing couplings. 

As mentioned above, some of these results can be explained by known symmetries but many can not. Thus this analysis throws into stark relief the key question that has already been raised by the known Yukawa coupling results. Are there extra, to date unknown, symmetries that we should be considering in heterotic compactifications, or do quasi-topological constraints lead to the vanishings of infinite numbers of couplings in four dimensional effective theories, without the need for a low energy symmetry? Either possibility is exciting and holds the promise to improve our understanding of the the type of field theories that result from heterotic compactifications.

\vspace{0.2cm}

The structure of the rest of this paper is as follows. In Section \ref{setup} we describe how higher order couplings in heterotic models are associated to obstructions to higher order holomorphic deformations of the gauge bundle. In particular we detail the conditions that are required for deformations to be unobstructed and thus for couplings to vanish. We consider both rank preserving and rank changing deformations of the gauge bundle. In Section \ref{vansec} we present one simple cause of vanishing higher order couplings in heterotic compactifications. Once more we investigate cases which both preserve and break the gauge group and give a number of explicit examples. In some of these examples the fact that the higher order couplings vanish can be confirmed independently. In others, the infinite number of zero couplings form a prediction rather than a confirmation of known results. In Section \ref{conc} we conclude and discuss future directions of research that will be pursued.

\section{Set up: couplings and the Kuranishi map} \label{setup}

We will begin our analysis with a discussion of higher order obstructions to flat directions, corresponding to superpotential terms, in heterotic compactifications \cite{kuranishi,Berglund:1995yu,Anderson:2011ty}. We will begin with the simplest case of deformations corresponding to bundle moduli, before moving on to more involved, gauge group breaking deformations in later subsections.

\subsection{Rank preserving deformations}

The contribution to the heterotic superpotential which is relevant for matter and bundle moduli couplings is as follows.
\begin{eqnarray} \label{W}
W=\int_X \omega_3^{\text{YM}} \wedge \Omega
\end{eqnarray}
Here $\omega_3^{\text{YM}}$ is the Chern-Simons form associated to the gauge fields. The superpotential (\ref{W}) is extremized by holomorphic gauge connections.
\begin{eqnarray} \label{fabzero}
F_{\overline{a}\overline{b}}=0
\end{eqnarray}
The superpotential (\ref{W}) is  perturbatively exact due to a non-renormalization theorem stemming from supersymmetry and axion shift symmetries. Thus, if we wish to look for flat directions of the superpotential, we need only look for deformations of the gauge connection, such that the equation (\ref{fabzero}) can be satisfied. As stated above, in this sub-section we will consider rank preserving deformations that correspond to bundle moduli. More general deformations will be considered in later sub-sections.

In analyzing heterotic theories, we perform a formal perturbative expansion on the gauge connection. This is associated to the usual infinite series of cubic and higher couplings of fields arising from (\ref{W}).
\begin{eqnarray} \label{exp}
A_{\overline{a}}=A^{(0)}_{\overline{a}}+A^{(1)}_{\overline{a}}+A^{(2)}_{\overline{a}}+\ldots
\end{eqnarray}
Demanding that (\ref{fabzero}) is preserved under first order variation simply returns the result that $A^{(1)}$ is $\overline{D}$ closed. Together with exact changes to $A^{(1)}$ being obtainable by gauge transformations, this leads to the standard result that the massless perturbations are counted by elements of cohomology. Higher order conditions arising from satisfying (\ref{fabzero}) can be easily computed and yield the following.
\begin{eqnarray} \label{higherorder}
f^{xyz} A^{(1)y}_{[\overline{a}}A^{(1)z}_{\overline{b}]} &=&- (\overline{D}_{[\overline{a}} A^{(2)}_{\overline{b}]})^x\\ \nonumber
2f^{xyz} A^{(1)y}_{[\overline{a}}A^{(2)z}_{\overline{b}]} &=&-(\overline{D}_{[\overline{a}} A^{(3)}_{\overline{b}]})^x \\ \nonumber
2f^{xyz} A^{(1)y}_{[\overline{a}}A^{(3)z}_{\overline{b}]} +f^{xyz} A^{(2)x}_{[\overline{a}}A^{(2)y}_{\overline{b}]} &=&-(\overline{D}_{[\overline{a}} A^{(4)}_{\overline{b}]})^x  \\ \nonumber
&\vdots&
\end{eqnarray}
In the above expressions the left hand side is always closed, if the lower orders are satisfied. This can be easily proven by direct computation and use of the Jacobi identity on the structure constants $f^{xyz}$. The potential constraint on a fluctuation comes from the possibility that the left hand sides of these expressions fail to be exact. Given this structure, the higher order obstructions to (\ref{fabzero}) vanish iff a fluctuation is in the kernel of the Kuranishi map \cite{kuranishi,Berglund:1995yu,Anderson:2011ty}.
\begin{eqnarray} \label{kurx}
H^1(\textnormal{End}_0(V)) \stackrel{k}{\longrightarrow} H^2(\textnormal{End}_0(V))
\end{eqnarray}
Here the map $k$ is defined, order by order, by the expressions in (\ref{higherorder}).

We can make contact between this notion of  unconstrained deformations/higher order obstructions and conventional notions of multi-particle couplings by parameterizing $A^{(1)}$ appropriately. Specifically, we can expand $A^{(1)}$ in a basis of appropriate bundle valued harmonic forms $\nu_i$ with coefficients $C^i$ which become the four-dimensional fields upon dimensional reduction.
\begin{eqnarray}
A^{(1)} = \sum_i C^i \nu_i
\end{eqnarray}
Looking at (\ref{higherorder}) we then see that the quantities $A^{(n)}$ are associated to $n$'th powers of the fields $C^i$. The first line in (\ref{higherorder}), for example, is then associated to cubic couplings. Looking at the coefficients of two specific $C^i$ for the two $A^{(1)}$ factors, we see that this expression becomes the statement that a map from two elements of $H^1(X,\text{End}_0(V))$ to $H^2(X, \text{End}_0(V))$,
\begin{eqnarray}
H^1(X,\text{End}_0(V))\times H^1(X,\text{End}_0(V)) \to H^2(X,\text{End}_0(V))\;,
\end{eqnarray}
has zero image. This is precisely the statement that all Yukawa couplings between those two fields and one other element of $H^1(X,\text{End}_0(V))$ vanish \cite{Distler:1987ee}. The later lines in (\ref{higherorder}) then correspond to the vanishing of couplings of higher orders.

\vspace{0.2cm}

This structure can also be seen in terms of the superpotential. The expansion (\ref{exp}) gives rise to the following contributions to $W$ order by order.
\begin{eqnarray}
W= W^{(3)} + W^{(4)} + \ldots \;,
\end{eqnarray}
where
\begin{eqnarray} \label{wexp1}
W^{(3)} &=& \frac{2}{3} \int_X f_{xyz}A^{(1)x} \wedge A^{(1)y} \wedge A^{(1)z} \wedge \Omega \\ \newline \label{wexp2}
W^{(4)}&=&3 \int_X f_{xyz} A^{(2)x} \wedge A^{(1)y} \wedge A^{(1)z} \wedge \Omega \\ \newline \nonumber
&\vdots& \;.
\end{eqnarray}
In deriving these expressions from (\ref{W}) and (\ref{exp}) we have used all lower order results in simplifying each given order, as well as some integration by parts. So for example, (\ref{wexp1}) uses that $F_{\overline{a}\overline{b}}=0$ and $\overline{D}A^{(1)}=0$. Equation (\ref{wexp2}) uses the same two conditions and the first condition in (\ref{higherorder}), and so on. It is easy to see, given the previous discussion, that the $W^{(n)}$ are contributions to the superpotential containing $n$'th powers of fields. For example, the $n=3$ term in (\ref{wexp1}) is recognizable as the expression for the Yukawa coupling \cite{Green:1987mn}. By contrast, one can see that $W^{(4)}$ in (\ref{wexp2}) gives rise to quartic couplings between the fields. More generally, $W^{(n)}$ corresponds to $n$'th order couplings between the fields of the low energy superpotential.  Thus, instead of following the analysis using (\ref{higherorder}) that we will largely pursue in this paper, one could analyze superpotential terms of this form and their derivatives order by order, so as to establish the existence of flat directions in four dimensional field space.

\vspace{0.2cm}

As a final comment, note that it is clear that, if a given $A^{(1)}$ is unobstructed at a given point in moduli space, and is associated with a particular direction in four dimensional field space denoted by $\phi$, then all couplings involving of any number of $\phi$ factors must vanish if the other fields are fixed to that point in field space. Thus contact between unobstructed deformations and vanishing couplings is easy to establish.

\subsection{Rank changing deformations}

We are often interested in rank changing deformations of the gauge bundle. This could either be because we are interested in Higgsing processes, or simply because we want to understand couplings involving charged matter. Let us look at two examples of how such deformations can be dealt with in this setting. We will start with an example involving chiral matter in this subsection, before moving on to the case of vector like degrees of freedom in the next.

\subsubsection{${\bf 27}$'s of $E_6$} \label{chiralsetup}

Again consider an expansion of the form (\ref{exp}) but where now the $A^{(1)}$ perturbation is associated to an element of $H^1(X,V)$ for an initially $SU(3)$ bundle. That is, we are considering perturbations associated to ${\bf 27}$ representations of $E_6$ in the four dimensional theory. This means we are regarding $A$ as a connection on an $E_8$ bundle and are allowing the group in which the expectation value of $A$ takes values to enlarge under deformation.

We can now consider an expansion such as (\ref{higherorder}) and reinterpret it in this context. The first line in (\ref{higherorder}) again corresponds to a Yukawa coupling vanishing, this time one associated to a map,
\begin{eqnarray}
H^1(X,V) \times H^1(X,V) \to H^2(X, \wedge^2V) = H^1(X, V)\;.
\end{eqnarray}
This Yukawa coupling is the standard ${\bf 27}^3$ one of the $E_6$ four dimensional GUT.

The next order is more complex. The constraint\footnote{Here, and in what follows, we will often use a more compact notation in the interests of readability, given that the gauge index structure in particular is becoming more involved.},
\begin{eqnarray}
2 A^{(1)} \wedge A^{(2)} = - \overline{D} A^{(3)}\;,
\end{eqnarray}
corresponds to a mapping \footnote{In fact, $A^{(2)}$ is essentially specified by $A^{(1)}$, so this can really be regarded as a mapping $H^1(X,V) \to H^2(X,{\cal O}) \oplus H^2(X,\text{End}_0(V))$},
\begin{eqnarray}
H^1(X,V) \times \Omega^1(X,\wedge^2 V) \to  H^2(X,{\cal O})\oplus H^2(X,\text{End}_0(V)) \;.
\end{eqnarray}
However, on a Calabi-Yau threefold $H^2(X,{\cal O})=0$ and so the requirement of exactness imposes no constraint in this component of the image. The other term in the image corresponds to a bundle modulus coupling of the form ${\bf 27}^3 {\bf 1}$. Given that we are only giving an expectation value to a ${\bf 27}$ field, it may seem strange to have such a possible obstruction. However, such a coupling would lead to a constraint, even if the bundle modulus was not given a vev away from its starting point, via the $\partial W=0$ constraint of a supersymmetric Minkowski vacuum where the derivative is taken with respect to the bundle modulus.

At next order we have,
\begin{eqnarray}
2 A^{(1)} \wedge A^{(3)} + A^{(2)} \wedge A^{(2)} = -\overline{D} A^{(4)}
\end{eqnarray}
The right hand side of this condition represents the zero element in $H^2(X,V)=H^1(X,V^{\vee})$. As with the previous order, this may at first seem strange as we were considering a ${\bf 27}$ deformation not a $\overline{{\bf 27}}$. However, this contribution corresponds to ${\bf 27}^4 \overline{{\bf 27}}$ couplings in the superpotential, and the derivative of this with respect to the $\overline{\bf 27}$ fields results in a constraint on the field space from the $\partial W=0$ condition which depends solely on the ${\bf 27}$ fields.

The pattern continues as the order increases, with all cases giving a contribution to a map into $H^1(X,V)$, $H^1(X, \text{End}_0(V))$ or $H^1(X,V^{\vee})$. Thus, an unobstructed deformation corresponds to one that is in the kernel of all of the following maps, where the maps are defined order by order by the above analysis.
\begin{eqnarray}
H^1(X,V) &\to& H^2(X,V^{\vee}) \\ \nonumber
H^1(X,V) &\to& H^2(X,\text{End}_0(V)) \\ \nonumber
H^1(X,V) &\to& H^2(X, V)
\end{eqnarray}

\subsubsection{${\bf 10}$'s of $SO(10)$} \label{so10setup}

For an application later in this paper, we will want to consider a perturbation to the gauge field corresponding to varying both bundle moduli and ${\bf 10}$'s of $SO(10)$ (for a bundle with initial structure group of $SU(4)$). At linear order, this corresponds to including both terms that are proportional to elements of $H^1(X, \text{End}_0(V))$ and terms that are proportional to elements of $H^1(X, \wedge^2 V)$. Thus, one is again considering $A$ as a connection on an $E_8$ bundle and is varying the connection so that the gauge field expectation values fill out a larger group after deformation.

Taking the vevs of both associate matter fields to be small so that a perturbative approach is appropriate, we can track the orders to which we are working by a bi-graded expansion, as opposed to an expansion indexed by a single integer $n$ as in the proceeding subsections. A term with superscript $(i,j)$ denotes a contribution that is order $i$ in the bundle modulus vev and order $j$ in that of the ${\bf 10}$.

\begin{eqnarray}
A = A^{(0,0)} + A^{(1,0)} + A^{(0,1)} + A^{(2,0)} + A^{(1,1)} + A^{(0,2)} + \ldots
\end{eqnarray}

One can then once more expand the gauge field strength order by order and demand that it vanish. The $(0,0)$, $(1,0)$ and $(0,1)$'th order results give what one would expect, the requirement of having a good vacuum, and the closedness of the linear perturbations of each type (so that they form elements of cohomology given their definition up to gauge transformations).

The degree $(1,1)$ contribution to $F_{\overline{a} \overline{b}}=0$ looks like the following.
\begin{eqnarray}\label{higherorder21}
A^{(1,0)} \wedge A^{(0,1)} + A^{(0,1)} \wedge A^{(1,0)} = - \overline{D} A^{(1,1)}
\end{eqnarray}
This is simply the statement that for the perturbation being considered to be unobstructed the $\bf{1} \bf{10} \bf{10} $ Yukawa coupling must vanish. That is, the map 
\begin{eqnarray} \label{genkur}
H^1(X, \text{End}_0(V)) \times H^1(X, \wedge^2 V) \to H^2(X, \wedge^2 V)\;
\end{eqnarray}
should have zero image.

At degree $(2,1)$ we have the following.
\begin{eqnarray} 
A^{(2,0)} \wedge A^{(0,1)} + A^{(0,1)} \wedge A^{(2,0)} + A^{(1,1)}\wedge A^{(1,0)} +A^{(1,0)} \wedge A^{(1,1)} = -\overline{D} A^{(2,1)}
\end{eqnarray}
Given the equations such as (\ref{higherorder}) and (\ref{higherorder21}) determining $A^{(2,0)}$ and $A^{(1,1)}$ the left hand side here is closed and is specified by our choice of $A^{(0,1)}$ and $A^{(1,0)}$. Thus this is again a map of the form (\ref{genkur}), taking into account second order effects in the bundle moduli deformation. This pattern then continues, with the degree $(n,1)$ perturbation to $F_{\overline{a} \overline{b}}=0$ corresponding to the order $n$ bundle modulus contribution to the ${\bf 10}$ mass term.
\begin{eqnarray} \label{ordn}
A^{(n,0)} \wedge A^{(0,1)} + A^{(0,1)} \wedge A^{(n,0)} + A^{(n-1,0)}\wedge A^{(1,1)} +A^{(1,1)} \wedge A^{(n-1,0)} +\ldots = -\overline{D} A^{(n,1)}
\end{eqnarray}
As before, the only constraint imposed on the fluctuations here, is that  the left hand side of expressions such as (\ref{ordn}) is exact. Hence, to all $n$, the constraints can be phrased as a Kuranishi style kernel associated to a mapping of the form (\ref{genkur}).

\section{Vanishings of  couplings} \label{vansec}

\subsection{Couplings of bundle moduli} \label{vanbm}

Consider a bundle $V$ over a Calabi-Yau threefold $X$ which is the pullback under the embedding map of a bundle $\hat{V}$ over some ambient space ${\cal A}$ in which $X$ is described as a complete intersection. Take a closed form $A^{(1)}$ which is associated to a class in $H^1(X,\text{End}_0(V))$ which lifts to a form $\hat{A}^{(1)}$ in $H^1({\cal A},\hat{V})$. Such perturbations to the gauge connection are referred to in the literature as type 1 \cite{Blesneag:2015pvz,Blesneag:2016yag}. It is then easy to show that if $\hat{A}^{(1)}$ is in the kernel of the Kuranishi map associated to $\hat{V} \to {\cal A}$ then $A^{(1)}$ is in the kernel of the Kuranishi map associated to $V\to X$. To show this observe that if we take the Koszul sequence, where ${\cal N}$ is the normal bundle to $X$ in ${\cal A}$, and tensor it up by $\hat{V}$ we obtain the following.
\begin{eqnarray}
\ldots \to {\cal N}^{\vee} \otimes \hat{V} \stackrel{p}{\longrightarrow} \hat{V} \stackrel{r}{\longrightarrow} V \to 0
\end{eqnarray}
The map $r$ is surjective and holomorphic. Now consider the analogue of (\ref{higherorder}), defining the Kuranishi map, for $\hat{V}$. Here we will only explicitly show one order but the argument is identical for any $n$. Assuming that $\hat{A}^{(1)}$ is unobstructed we have,
\begin{eqnarray}
2 f^{xyz}\hat{A}^{(1)y} \wedge \hat{A}^{(2)z} = -\left(\overline{D} \hat{A}^{(3)}\right)^x\;.
\end{eqnarray}
Applying the restriction map and using the holomorphy of $r$, we then obtain the following.
\begin{eqnarray}
&&r \left(2 f^{xyz} \hat{A}^{(1)y} \wedge \hat{A}^{(2)z}\right) = -r\left(\overline{D} \hat{A}^{(3)}\right)^x \\ \nonumber
&\Rightarrow& 2 f^{xyz} r(\hat{A}^{(1)y}) \wedge r(\hat{A}^{(2)z}) = -r\left(\overline{D} \hat{A}^{(3)}\right)^x \\ 
&\Rightarrow& 2 f^{xyz} A^{(1)y}\wedge  A^{(2)z} = -\left(\overline{D} \,r(\hat{A}^{(3)})\right)^x \\ 
\end{eqnarray}
We see that the solution to the constraint equation on ${\cal A}$ implies the solution to the constraint equation on $X$ where we take $A^{(3)} =r(\hat{A}^{(3)})$.

The question now becomes, when will $\hat{A}^{(1)}$ be unobstructed. There are many fashions in which this can happen, but the most straightforward is simply if $H^2({\cal A},\text{End}_0(V))=0$\footnote{Note that the analogous condition can not arise on $X$. There, Serre duality implies that $H^2(X,\text{End}_0(V)) = H^1(X,\text{End}_0(V))$ and thus the $H^2$ vanishing would mean that we didn't have any bundle moduli in the first place.}. This leads us to the following statement.

\vspace{0.2cm}
\noindent
{\bf Lemma:} {\it Consider a case where a bundle $V \to X$ descends from a bundle $\hat{V} \to {\cal A}$ where $X$ is a complete intersection in ${\cal A}$. If a bundle modulus is type 1, that is if the associated element of $H^1(X, \text{End}_0(V))$ descends from an element of $H^1({\cal A},\text{End}_0(\hat{V}))$, and if $H^2({\cal A}, \text{End}_0(\hat{V}))=0$ then that bundle modulus is unobstructed, and all perturbative couplings, of any order, which could obstruct that perturbation vanish.}
\vspace{0.2cm}

Note that this lemma is particularly powerful for $X$'s which are hypersurfaces inside the ambient ${\cal A}$. In this case, from the long exact sequence associated to,
\begin{eqnarray} \label{kos}
0 \to {\cal N}^{\vee} \otimes \text{End}_0(\hat{V}) \stackrel{p}{\longrightarrow} \text{End}_0(\hat{V}) \stackrel{r}{\longrightarrow} \text{End}_0(V)\to 0,
\end{eqnarray}
we have that,
\begin{eqnarray}
H^1(X,\text{End}_0(V)) =&& \text{Ker}(H^2({\cal A},{\cal N}^{\vee} \otimes \text{End}_0(\hat{V})) \to H^2({\cal A},\text{End}_0(\hat{V})))\\ \nonumber
&&\;\;\;\;\;\;\;\;\;\;\;\;\;\;\oplus \\ \nonumber &&\text{Coker}(H^1({\cal A},{\cal N}^{\vee} \otimes \text{End}_0(\hat{V}))\to H^1({\cal A}, \text{End}_0(\hat{V})))\;.
\end{eqnarray}
Using Serre duality on the ambient space we have that $h^2({\cal A},{\cal N}^{\vee} \otimes \text{End}_0(\hat{V}))= h^2({\cal A},\text{End}_0(\hat{V}))$, where we have used the fact that $\text{End}_0(\hat{V})$ is self dual. Thus, the statement that $H^2({\cal A},\text{End}_0(\hat{V}))=0$ implies that {\it all} bundle moduli are automatically of type 1. We are therefore lead to the following corollary.

\vspace{0.2cm}
\noindent
{\bf Corollary:} {\it If a bundle $V$ over a Calabi-Yau threefold $X$ lifts to a bundle $\hat{V}$ over an ambient space ${\cal A}$  in which $X$ is a hypersurface and if $H^2({\cal A},\text{End}_0(\hat{V}))=0$ then all perturbative couplings of all orders between bundle moduli vanish.}
\vspace{0.2cm}

Let us now give a concrete example where we can see that the above vanishing theorem reproduces expected results.

\subsubsection{An example} \label{eg1}
Consider the following Calabi-Yau threefold $X$ and bundle $V$.
\begin{eqnarray} \label{egX1}
X= \left[\begin{array}{c|cc} \mathbb{P}^1&1&1 \\ \mathbb{P}^4&1&4\end{array}\right]
\end{eqnarray}
\begin{eqnarray} \label{egV1}
0\to V\to {\cal O}_X(1,0)^4 \oplus {\cal O}_X(0,1)^2 \to {\cal O}_X(2,1)^2 \to 0
\end{eqnarray}
Naively this system presents the type of puzzle that was discussed in the introduction. Computing the bundle moduli of this system, one finds that they are purely associated to monad maps. Writing $B_X=  {\cal O}_X(1,0)^3 \oplus {\cal O}_X(0,1)^3$ and $C_X={\cal O}_X(1,1) \oplus {\cal O}_X(2,2)$ we find $H^1(X,\text{End}_0(V)) \subset H^0(X,B_X^{\vee}\otimes C_X)$. This tells us that, at least infinitesimally, the choices of monad map cover the bundle moduli space. Furthermore, all variations of monad map that are sufficiently generic, whether infinitesimal or not, lead to well defined holomorphic stable bundles. Given this, one might naively expect that all of the infinite possible set of self couplings between all of the bundle moduli vanish.

This is somewhat confusing however, at least at first glance. From a field theory perspective, the bundle moduli are singlets in the four-dimensional theory. Therefore, there is no obvious reason why such couplings would vanish. From a mathematical perspective, the situation is also not clear, because $h^2(X,\text{End}_0(V))\neq0$, and thus there is no obvious reason to expect the Kuranishi map (\ref{kurx}) to have trivial image.

The lemma of this section can be used to explain this result. The bundle (\ref{egV1}) does {\it not} lift to a bundle on the ambient space $\mathbb{P}^1\times \mathbb{P}^4$. It does, however, lift to a bundle on an ambient space 
\begin{eqnarray}\label{Aeg1}
{\cal A}= \left[\begin{array}{c|c} \mathbb{P}^1 & 1 \\ \mathbb{P}^4 &1 \end{array} \right]
\end{eqnarray}
We can thus regard $X$ as a $(1,4)$ hypersurface in this ambient space and define a bundle $\hat{V}$ which pulls back to $V$ by a sequence identical to (\ref{egV1}), with the exception that the line bundles are over ${\cal A}$ as in (\ref{Aeg1}). In terms of this description, one can compute the cohomologies associated to elements of the Koszul sequence and show that the bundle moduli are of type 1 and $H^2({\cal A} ,\text{End}_0(\hat{V}))=0$. In more detail, here is the Koszul sequence with the dimensions of the cohomology groups $h^i$ written underneath the associated bundles, starting with $i=0$ and increasing down the columns.
\begin{eqnarray} \label{bunmodh2}
\begin{array}{ccccccc} 
0 \to & {\cal N}^{\vee} \otimes \text{End}_0 (\hat{V})& \to &\text{End}_0 (\hat{V}) &\to &\text{End}_0(V) &\to 0 \\ 
&0&&0&&0& \\
&0&&48&&48& \\
&0&&0&&48& \\
&48&&0&&0& \\
&0&&0&&& 
\end{array}
\end{eqnarray}
Thus the corollary from earlier in this section applies and all perturbative couplings between the bundle moduli of this system vanish, explaining the observed structure in the compactification.

\subsubsection{Non-genericity and other resolutions}

One may ask what other ambient space structures can lead to trivial Kuranishi obstructions. To answer such a question we are interested in the opposite line of reasoning to that given above. Instead of assuming that the Kuranishi obstruction is trivial on the ambient space and discussing consequences on the Calabi-Yau threefold, we should assume that the Kuranishi obstruction is trivial on $X$ and see what that implies on the ambient space ${\cal A}$.

We restrict our attention to a case where the Koszul sequence (\ref{kos}) holds, and where the bundle moduli of interest are type 1. Considering (\ref{higherorder}) we have, at all orders, equations of the form $\text{L.H.S}+(\overline{D} A^{(n)})=0$, where we have abbreviated the left hand side of each of the equations, whose exact form we will not need in the following. The lift $\hat{A}^{(k)}$ of the $A^{(k)}$'s then obeys an equation of the form $r(\text{L.H.S}+(\overline{D} \hat{A}^{(n)}))=0$. The exactness of the Koszul sequence then implies that $\text{L.H.S}+(\overline{D} \hat{A}^{(n)})=p \hat{\omega}^{(n)}$ where $\hat{\omega}^{(n)}\in H^2({\cal A},{\cal N}^{\vee} \otimes \text{End}_0(\hat{V}))$. The closedness of $\hat{\omega}^{(n)}$ follows from the use of any lower order results, and the Jacobi Identity, and the assumption that $\hat{A}^{(1)}$ is closed, just as in Section \ref{setup}. Given this, the question of what ambient space structure causes the obstruction to vanish on $X$ reduces to a question about $p\hat{\omega}$. If $p\hat{\omega}$ is exact, then one can redefine $\hat{A}^{(n)}\to \hat{A}^{(n)}- \hat{\Gamma}^{(n)} $ where $p\hat{\omega}^{(n)} = \overline{D} \hat{\Gamma}^{(n)}$ and we have the previous case discussed where an unobstructed deformation on the ambient space is reducing to an unobstructed deformation on the Calabi-Yau threefold. In particular, in the case of the vanishings we have been discussing $p\hat{\omega}$ is exact because it is an element of the group $H^2({\cal A} ,\text{End}_0(\hat{V}))$ which is taken to be trivial. If $p\hat{\omega}$ is not exact then we have a situation where an obstructed deformation on the ambient space restricts to an unobstructed one on the Calabi-Yau threefold.

One could imagine non-obstructed deformations on $X$ lifting to non-obstructed deformations on ${\cal A}$ when $H^2({\cal A} ,\text{End}_0(\hat{V}))$ is not trivial. For example, if $h^2({\cal A},{\cal N}^{\vee}\otimes \text{End}_0(\hat{V}))=0$ then $\hat{\omega}^{(n)}$ is automatically exact and therefore so is $p\hat{\omega}^{(n)}$. In codimension one, this is no different from the case we have been considering because, by Serre duality, $h^2({\cal A},{\cal N}^{\vee}\otimes \text{End}_0(\hat{V}))=h^2({\cal A},\text{End}_0(\hat{V}))$. More generally however, in higher codimension, the two cohomologies are not equal in general, and so different cases are possible. 

\vspace{0.2cm}

The above analysis implies that it is not always the case that vanishing couplings on $X$ will be associated with unobstructed deformations on ${\cal A}$. As such we should consider more general structures than the Koszul resolution in trying to explain these phenomena. It is important to note that, while the lemma given in this section has been stated in terms of a Koszul resolution for a complete intersection Calabi-Yau threefold, very little of this specific structure is needed in our arguments. For example, instead of a complete intersection in ${\cal A}$,  $X$ could be any subvariety such that there exists a surjective holomorphic restriction map from $\hat{V}$ to $V$. Even more generally, all we require is any surjective holomorphic map $r$ of any type between a bundle $\hat{V}$ on some space ${\cal A}$ (which could be $X$ for example) and a bundle $V$ on $X$. Examples of the use of a different resolutions from the Koszul sequence are something we will return to in future work.

\subsection{Couplngs of chiral matter}

The arguments of Section \ref{vanbm} can be essentially repeated for the case where we consider chiral matter of $E_6$ as in Section \ref{chiralsetup}. In the case where we are considering couplings that could obstruct a ${\bf 27}$ field flat direction, the relevant maps whose image must be trivial if those couplings are to vanish are as follows.
\begin{eqnarray} \label{chiralrecap}
H^1(X,V) &\to& H^2(X, V^{\vee}) \\ \nonumber
H^1(X,V) &\to& H^2(X,\text{End}_0(V)) \\ \nonumber
H^1(X,V) &\to& H^2(X, V)\;.
\end{eqnarray}
If we assume that $A^{(1)}$ is of type 1, then, by the same reasoning as given in Section \ref{vanbm}, the deformations are unobstructed if $H^2({\cal A}, \hat{V}^{\vee})=H^2({\cal A},\hat{V})=H^2({\cal A},\text{End}_0(\hat{V}))=0$.  The first line in (\ref{chiralrecap}) corresponds to ${\bf 27}^{3m}$ couplings, the second line to ${\bf 27}^{3m} {\bf 1}$ couplings (where the singlet is a bundle modulus) and the last line to ${\bf 27}^{3m+1} \overline{\bf 27}$ couplings where $m \in \mathbb{Z}$. Gathering this information together we have the following.

\vspace{0.2cm}
\noindent
{\bf Lemma:} {\it Consider a case where an $SU(3)$ bundle $V \to X$ descends from a bundle $\hat{V} \to {\cal A}$ where $X$ is a complete intersection in ${\cal A}$. If a ${\bf 27}$ matter field is type 1, that is if the associated element of $H^1(X, V)$ descends from an element of $H^1({\cal A},\hat{V})$, and if $H^2({\cal A}, \hat{V})=H^2({\cal A}, \hat{V}^{\vee})=H^2({\cal A},\text{End}_0(\hat{V}))=0$ then that ${\bf 27}$ field is unobstructed, and all perturbative couplings, of any order, which could obstruct that perturbation vanish.}
\vspace{0.2cm}

\subsubsection{Two examples}

Let us start with an example of vanishings of infinite sets of couplings between chiral matter fields where we can verify the result. Consider the Calabi-Yau threefold,
\begin{eqnarray} \label{the24}
X= \left[ \begin{array}{c|c} \mathbb{P}^1 &2 \\ \mathbb{P}^3 &4\end{array} \right]\;,
\end{eqnarray}
with the following sum of line bundles over it.
\begin{eqnarray} \label{linesum}
V= {\cal O}_X(-2,1)^{\oplus 2} \oplus {\cal O}_X(4,-2)
\end{eqnarray}
This $SU(3)$ bundle (strictly $S(U(1)^3)$ as we will return to shortly) lifts to a bundle on the ambient space, and we have the following Koszul sequences\footnote{Note that is is perfectly consistent for a poly-stable sum of line bundles to have non-vanishing global sections and top cohomology of $V\otimes V^{\vee}$.} for $V$, $V^{\vee}$ and $V\otimes V^{\vee}$.
\begin{eqnarray}
\begin{array}{ccccccc}
0 \to &{\cal N}^{\vee} \otimes \hat{V}& \to &\hat{V} &\to& V&\to 0 \\ 
&0&&0&&0& \\
&0&&8&&8& \\
&0&&0&&30& \\
&30&&0&&0&\\
&0&&0&&&
\end{array}
\end{eqnarray}
\begin{eqnarray}
\begin{array}{ccccccc}
0 \to &{\cal N}^{\vee} \otimes \hat{V}^{\vee}& \to &\hat{V}^{\vee} &\to& V^{\vee}&\to 0 \\ 
&0&&0&&0& \\
&0&&30&&30& \\
&0&&0&&8& \\
&8&&0&&0&\\
&0&&0&&&
\end{array}
\end{eqnarray}
\begin{eqnarray}
\begin{array}{ccccccc}
0 \to &{\cal N}^{\vee} \otimes \hat{V}\otimes \hat{V}^{\vee}& \to &\hat{V}\otimes \hat{V}^{\vee} &\to& V \otimes V^{\vee}&\to 0 \\ 
&0&&5&&5& \\
&0&&100&&100& \\
&0&&0&&	100& \\
&100&&0&&5&\\
&5&&0&&&
\end{array}
\end{eqnarray}
We see that all fields are of type 1 and that $H^2({\it A},\hat{V}) =H^2({\cal A},\hat{V}^{\vee})=H^2({\cal A},\text{End}_0(\hat{V}))=0$. Thus, our analysis should hold and, for example, all of the ${\bf 27}$ matter fields should be unobstructed.

As mentioned above, the structure group of (\ref{linesum}) is actually $S(U(1)^3)$ rather than $SU(3)$. This means that there are two extra Green-Schwarz anomalous $U(1)$ symmetries in the low energy effective theory. We can now check that, in this case, all of the couplings that must vanish in order for the ${\bf 27}$ directions in field space to be unobstructed are forbidden by these additional symmetries.

In the notation of \cite{Anderson:2011ns,Anderson:2012yf}, charges under the $U(1)^2$ factor of the gauge group can be represented as a vector of 3 entries where two vectors are identified as $S(U(1)^3)$ representations if ${\bf q}-\tilde{\bf q} \in \mathbb{Z} (1,1,1)$. Following the analysis of \cite{Anderson:2011ns,Anderson:2012yf}, we have four ${\bf 27}_{1,0,0}$ four ${\bf 27}_{0,1,0}$ and 30 $\overline{\bf 27}_{0,0,1}$ fields in this example. So all couplings purely between ${\bf 27}$ or purely between ${\overline{\bf 27}}$ are forbidden by the symmetry. It is permitted by the symmetry to have mixed ${\bf 27}$, $\overline{\bf27}$ couplings. However, only those of the form ${\bf 27}_{1,0,0}^{3n}{\bf 27}_{0,1,0}^{3n}\overline{\bf 27}^{3n}_{0,0,1}$ where $n\in \mathbb{Z}$ are consistent with the full $E_6\times S(U(1)^3)$ gauge group. The vanishing of these couplings are not required for the above ``no-obstruction" result to hold. One can include the bundle moduli in the couplings as well. These have charges $(1,0,-1)$ or $(0,1,-1)$. However, even with their presence, the only allowed couplings are those that have more than one field that is not a ${\bf 27}$ representation. Thus we find that in this case our analysis does indeed give a correct result and this can be confirmed by symmetry considerations.

\vspace{0.2cm}

We would like to emphasize that, while we have chosen the above example because the infinite set of couplings that we predict to vanish can be confirmed to do so by a known symmetry, it is certainly not known to be the case that every set of vanishings our analysis predicts can be explained in this manner. For example, on the tetra-quadric,
\begin{eqnarray}
X= \left[ \begin{array}{c|c} \mathbb{P}^1 & 2 \\\mathbb{P}^1 & 2 \\\mathbb{P}^1 & 2 \\\mathbb{P}^1 & 2 \end{array} \right]\;,
\end{eqnarray}
consider the following monad bundle.
\begin{eqnarray}
0 \to V \to {\cal O}_X(1,0,0,0)^{\oplus 2} \oplus {\cal O}_X(0,1,0,0) \oplus {\cal O}_X(0,0,1,0) \to {\cal O}_X(2,1,1,0) \to 0
\end{eqnarray}
This bundle descends from a bundle $\hat{V}$ defined by an identical monad sequence on the ambient space. The Koszul sequences for this bundle, its dual and its traceless endomorphisms look as follows.
\begin{eqnarray}
\begin{array}{ccccccc}
0 \to &{\cal N}^{\vee} \otimes \hat{V}& \to &\hat{V} &\to& V&\to 0 \\ 
&0&&0&&0& \\
&0&&4&&4& \\
&0&&0&&0& \\
&0&&0&&0&\\
&0&&0&&&
\end{array}
\end{eqnarray}
\begin{eqnarray}
\begin{array}{ccccccc}
0 \to &{\cal N}^{\vee} \otimes \hat{V}^{\vee}& \to &\hat{V}^{\vee} &\to& V^{\vee}&\to 0 \\ 
&0&&0&&0& \\
&0&&0&&0& \\
&0&&0&&4& \\
&4&&0&&0&\\
&0&&0&&&
\end{array}
\end{eqnarray}
\begin{eqnarray}
\begin{array}{ccccccc}
0 \to &{\cal N}^{\vee} \otimes \text{End}_0(\hat{V})& \to &\text{End}_0(\hat{V}) &\to& \text{End}_0(V)&\to 0 \\ 
&0&&0&&0& \\
&0&&21&&21& \\
&0&&0&&21& \\
&21&&0&&0&\\
&0&&0&&&
\end{array}
\end{eqnarray}
In particular we see that all matter fields are of type 1 and $H^2({\cal A},\hat{V})=H^2({\cal A},\hat{V}^{\vee})=H^2({\cal A},\text{End}_0(\hat{V}))=0$ and thus our vanishing coupling analysis holds. In fact, given that this result holds at a generic point in bundle moduli space, in this case we would not expect the {\bf 27} matter fields to appear in the perturbative superpotential at all.

In this case there is no obvious symmetry in the four dimensional effective theory that would forbid such couplings. Just as has been seen in the Yukawa coupling literature, the structure of the embedding of the geometry of the compactification in an ambient space is leading to a vanishing of couplings that is mysterious from the point of view of our current understanding of the four dimensional physics. One can perform a consistency check and compute the Yukawa couplings using standard techniques \cite{Distler:1987gg,Distler:1987ee,Anderson:2010vdj}, finding that these are indeed vanishing. However, the results we have garnered here go way beyond merely cubic couplings.

\subsection{Couplings of bundle moduli and vector like matter}

The arguments of Section \ref{vanbm} again repeat, essentially unmodified, for the situation considered in Section \ref{so10setup}. In this case of couplings between bundle moduli and ${\bf 10}$'s of $SO(10)$ which can give rise to mass terms for the vector like matter, our map is of the form
\begin{eqnarray}
H^1(X, \text{End}_0(V)) \times H^1(X, \wedge^2 V) \to H^2(X, \wedge^2 V)\;.
\end{eqnarray}
If we assume that both $A^{(1,0)}\in H^1(X, \text{End}_0(V))$ and $A^{(0,1)}\in H^1(X, \wedge^2 V) $ are of type 1, then, by exactly the same line of reasoning as in Section \ref{vanbm}, the deformations are unobstructed if $H^2({\cal A}, \wedge^2 \hat{V})=0$. In such a situation, all of the couplings that could obstruct such a deformation vanish, and in particular the contribution of all powers of the associated bundle moduli to the mass terms for the ${\bf 10}$'s vanish. Thus, the vector like matter would be expected to remain massless when we give vevs to those bundle moduli.

\vspace{0.2cm}
\noindent
{\bf Lemma:} {\it Consider a case where a bundle $V \to X$ descends from a bundle $\hat{V} \to {\cal A}$ where $X$ is a complete intersection in ${\cal A}$. Take a bundle modulus that is type 1, that is the associated element of $H^1(X, \text{End}_0(V))$ descends from an element of $H^1({\cal A},\text{End}_0(\hat{V}))$, and a ${\bf 10}$ of $SO(10)$ which is also type 1, with the associated element of $H^1(X, \wedge^2V)$ descending from an element of $H^1({\cal A},\wedge^2\hat{V})$. Then, if $H^2({\cal A},\wedge^2\hat{V})=0$, all couplings between  any number of those bundle moduli and two of those ${\bf 10}$'s vanish. Thus the vector like matter is perturbatively massless for all values of those bundle moduli.}
\vspace{0.2cm}

\subsubsection{An example}

Returning to the example of Section \ref{eg1} we find the following Koszul sequence associated to $\wedge^2 V$, again taking  (\ref{Aeg1}) as the ambient space. 
\begin{eqnarray} \label{grid1}
\begin{array}{ccccccc} 
0 \to & {\cal N}^{\vee} \otimes \wedge^2\hat{V}& \to &\wedge^2 \hat{V} &\to &\wedge^2 V &\to 0 \\ 
&0&&0&&0& \\
&0&&6&&6& \\
&0&&0&&6& \\
&6&&0&&0& \\
&0&&0&&& 
\end{array}
\end{eqnarray}
We see from this data that all of the perturbations $A^{(0,1)}$ are of type 1 since all six such degrees of freedom descend from $H^1({\cal A},\wedge^2\hat{V})$. We have already seen in Section \ref{eg1} that all of the bundle moduli are of type 1. Finally, we also see from (\ref{grid1}) that $H^2({\cal A}, \wedge^2 \hat{V})=0$. Thus, from the above analysis we expect all of the bundle moduli couplings to two ${\bf 10}$ fields to vanish. This expectation is corroborated by the cohomology computation which shows that all six families of vector like matter remain massless everywhere in bundle moduli space.

\section{Conclusions and future directions} \label{conc}

In this paper we have seen that just a few vanishing ambient space cohomologies in the description of a heterotic compactification can lead to exactly zero n-point couplings between certain four-dimensional fields for all n. Indeed, more generally, one would expect that vanishing cohomologies in a variety of different structures associated to the compactification can lead to such sparse interaction structure.

Vanishing higher order couplings of this type do not seem to be rare: the examples given in this paper were not difficult to construct. It would be interesting to see exactly how ubiquitous the applicability of these results is in standard approaches to heterotic model building. This would be a generalization of the type of investigation that was performed for the case of vanishing Yukawa couplings in \cite{Gray:2019tzn}.

It should be emphasized that it does not need to be the case that the Calabi-Yau and bundle of interest in a heterotic compactification are being described in a manner that enjoys the structure required in order for our results to hold. Rather, it is only necessary that such a structure {\it exists}. Given the fact that a given Calabi-Yau and bundle can be described in a myriad of different ways the constraints on interactions we have been discussing can be more restrictive than one would initially think \cite{Anderson:2022kgk}.

There is a key question that, while already raised by vanishing results in the Yukawa coupling literature \cite{Blesneag:2015pvz,Blesneag:2016yag,Anderson:2021unr},  is thrown into stark relief by the types of structure seen in this paper. What is the underlying physics that is causing these couplings to be precisely zero? There are essentially two possiblities. Either there is an important symmetry present in the four-dimensional effective theory of heterotic compactifications that has so far not been elucidated. Or these coupling vanish purely due to quasi-topological constraints that can not be explained by four dimensional symmetries. The former case would clearly be important in our understanding of these compactifications. The latter would mean that we were observing structure in the four-dimensional field theory that is surprising from a purely field-theoretic point of view and which therefore provides something of a signature of higher dimensional physics. Clearly, either of these possibilities is exciting, and we will be actively working to try and isolate which possibility is realized.

\vspace{0.2cm}

Technically there are a number of further results which one could try to establish. It would be nice to prove vanishing results associated to fields of higher type, as has been achieved in the Yukawa coupling case \cite{Blesneag:2015pvz,Blesneag:2016yag,Anderson:2021unr}. It is also very likely that progress can be made in considering other resolutions and structures of the compactification beyond those considered here. For example, one might expect that similar results could be achieved by making use of various fibration structures \cite{Braun:2006me}.  One of the biggest restrictions of the analysis in this paper is that $\hat{V}$ must be a bundle in order for our methods to hold. Clearly, it would be very desirable to relax this constraint, and include the possibility of sheaves that restrict to $V$, as was done in \cite{Anderson:2021unr} for the Yukawa coupling case. Finally, it would be interesting to fold the complex structure moduli into our analysis in an intrinsic way, utilizing a higher order version of the Atiyah style analysis found in \cite{Atiyah,Anderson:2010mh,Anderson:2011ty}. This last direction is one which we are actively pursuing (see \cite{delaOssa:2015maa,McOrist:2019mxh} for related work).

\section*{Acknowledgements}

James Gray is supported in part by NSF grant PHY-2310588. This research was supported in part by grant NSF PHY-2309135 to the Kavli Institute for Theoretical Physics (KITP). The author would like to thank the KITP, where the majority of this work was carried out, for it's hospitality and stimulating environment. 

\appendix

\section{Superpotential based proof} \label{appendix1}

In this appendix we give a simple proof of the vanishing of one higher order interaction in a manner that echos the derivations that have been carried out in parts of the literature for vanishing Yukawa couplings \cite{Blesneag:2015pvz,Blesneag:2016yag}. This type of proof does not seem to be the most efficient way to proceed in this context, and thus we simply give an example of a four-point coupling vanishing in a codimension one Calabi-Yau for illustrative purposes. This derivation can easily be generalized to include higher codimensions and higher couplings.

Let us examine a four field superpotential coupling between bundle moduli.
\begin{eqnarray} \label{vanone}
 W^{(4)}=2\int_X f_{xyz} A^{(1)x} \wedge A^{(1)y} \wedge A^{(2)z} \wedge \Omega
\end{eqnarray}
Consider the case where the Calabi-Yau manifold $X$ is described as a hypersurface inside some ambient space ${\cal A}$, and the bundle on the Calabi-Yau $V$ is the restriction of a bundle on the ambient space $\hat{V}$. We can then lift (\ref{vanone}) to an integral over ${\cal A}$ simply by making use of an appropriate delta function.
\begin{eqnarray}
W^{(4)}= 2\int_{\cal A} f_{xyz} \hat{A}^{(1)x} \wedge \hat{A}^{(1)y} \wedge \hat{A}^{(2)z} \wedge \hat{\Omega} \wedge \delta^2(p) dp \wedge d \overline{p}
\end{eqnarray}
In this expression, hatted quantities are ambient space forms which restrict to the associated un-hatted quantity on $X$ and $p$ is the defining relation of the Calabi-Yau threefold. Now, following \cite{Blesneag:2015pvz,Blesneag:2016yag}, define $\hat{\Omega}\wedge dp= \mu$ and use identity 
\begin{eqnarray} \label{ident1}
\delta^2(p) d\overline{p}=\frac{1}{\pi} \overline{\partial} \frac{1}{p}
\end{eqnarray}
to rewrite the integrand.
\begin{eqnarray}
W^{(4)}=\frac{2}{\pi}\int_{\cal A} f_{xyz} \hat{A}^{(1)x} \wedge \hat{A}^{(1)y} \wedge \hat{A}^{(2)z} \wedge \mu \wedge \overline{\partial} \frac{1}{p}
\end{eqnarray}
Next we integrate by parts and consider a case where $\overline{D} \hat{A}^{(1)x}=0$. In the language of the main text, we are considering a case where the cohomology element associated to $A^{(1)x}$ is of type 1.  We then obtain the following (working in ``math gauge" where the anti-holomorphic part of the zeroth order gauge field vanishes). 
\begin{eqnarray}
W^{(4)}=-\frac{2}{\pi} \int_{\cal A} f_{xyz} \hat{A}^{(1)x} \wedge\hat{A}^{(1)y} \wedge  \overline{D} ( \hat{A}^{(2)z}) \wedge \mu \frac{1}{p}
\end{eqnarray}
Let us denote the restriction map from ${\cal A}$ to $X$ (the pullback under the embedding) by $r$. Since $r$ is a holomorphic map we have that $r(2 (\overline{D} \hat{A}^{(2)})^x+f^{xyz} \hat{A}^{(1)y} \hat{A}^{(1)z} )=2 (\overline{D} {A}^{(2)})^x+f^{xyz} {A}^{(1)y} {A}^{(1)z} =0$, where we have assumed that the Yukawa couplings vanish so that $A^{(2)}$ is defined. Given the exactness of the Koszul sequence,
\begin{eqnarray}
0 \to {\cal N}^{\vee} \otimes \text{End}_0(\hat{V}) \stackrel{p}{\longrightarrow} \text{End}_0(\hat{V}) \stackrel{r}{\longrightarrow} \text{End}_0(V) \to 0 \;,
\end{eqnarray}
we therefore know that $2 (\overline{D} \hat{A}^{(2)})^x+f^{xyz} \hat{A}^{(1)y} \hat{A}^{(1)z} = p \omega^{(2)x}$ for some $\omega^{(2)x} \in H^2({\cal A},{\cal N}^{\vee} \otimes \text{End}_0(\hat{V}))$. Note that $\omega^{(2)x}$ is indeed closed because of the type 1 assumption on $A^{(1)}$ implying that $\hat{A}^{(1)}$ is closed. We will further consider a case where $H^2({\cal A}, \text{End}_0(\hat{V}))=0$. By Serre duality this then implies that $H^2({\cal A},{\cal N}^{\vee} \otimes \text{End}_0(\hat{V}))=0$ and so $\omega^{(2)x}$ is exact. Setting $\omega^{(2)x}= \overline{D} \Gamma^{(2)x}$ we are then free to redefine $\hat{A}^{(2)} \to \hat{A}^{(2)} - p \Gamma^{(2)}$. This new choice of $\hat{A}^{(2)}$ still satisfies $r(\hat{A}^{(2)})=A^{(2)}$ and so is also a valid choice of lift of that quantity to the ambient space. In addition, however, we now also have that $2 (\overline{D} \hat{A}^{(2)})^x+f^{xyz} \hat{A}^{(1)y} \hat{A}^{(1)z} =0$. Essentially, the Kuranishi conditions for $V$ over $X$ are lifting to the Kuranishi conditions for $\hat{V}$ over ${\cal A}$.

Using the analysis of the previous paragraph, our coupling can now be written as follows.
\begin{eqnarray}
W^{(4)}=-\frac{2}{\pi} \int_{\cal A} f_{xyz} \hat{A}^{(1)x} \wedge \hat{A}^{(1)y} \wedge\left(-\frac{1}{2} f^{zy'z'} \hat{A}^{(1)y'}\wedge \hat{A}^{(1)z'} \right) \wedge  \mu \frac{1}{p} 
\end{eqnarray}
The same computation that showed that the left hand sides of (\ref{higherorder}) are closed immediately shows that this quantity vanishes, as can be seen by direct calculation and use of the Jacobi identity. We are therefore lead to the following conclusion:

\vspace{0.2cm}
\noindent
{\it Consider  a bundle $V$ on a Calabi-Yau $X$ which lifts to a bundle $\hat{V}$ on an ambient space ${\cal A}$ in which $X$ is a hypersurface. Then a perturbation in field space which is of type 1, is unconstrained by fourth order couplings in the superpotential if $H^2({\cal A},\text{End}_0(\hat{V}))=0$.}
\vspace{0.2cm}

Obviously, this is merely a sub-result of the vanishing conditions given in the main text.


\end{document}